\documentclass[usenatbib]{mnras}
\usepackage{graphicx}
\usepackage{color}
\usepackage{amssymb}
\usepackage{multirow}
\usepackage{amsmath}
\usepackage{url}
\usepackage{txfonts}
\usepackage{textcomp}

\def\Msun{M$_{\odot}$}

\title[SZ observation of CIZA J2242+5301]{AMI SZ observation of galaxy-cluster merger CIZA J2242+5301: perpendicular flows of gas and dark matter}
\author[Rumsey et~al.]{Clare Rumsey,$^{1}$\thanks
{Email:cr461@mrao.cam.ac.uk} Yvette C. Perrott$^{1}$, Malak Olamaie$^{1,2}$, Richard D. E. Saunders$^{1,3}$,
\newauthor Michael P. Hobson$^1$, Andra Stroe$^{4}$, Michel P. Schammel$^{1}$, Keith J. B. Grainge$^{5}$ \vspace{0.3cm}\\
$^{1}$ Astrophysics Group, Cavendish Laboratory, 19 J. J. 
Thomson Avenue, Cambridge, CB3 0HE\\ 
$^{2}$Imperial Centre for Inference and Cosmology(ICIC), Imperial College,Prince
Consort Road, London SW7 2AZ \\
$^{3}$ Kavli Institute for Cosmology Cambridge, Madingley 
Road,Cambridge, CB3 0HA\\
$^{4}$European Southern Observatory, Karl-Schwarzschild-Str. 2, 
85748, Garching, Germany\\
$^{5}$ Jodrell Bank Centre for Astrophysics, 
School of Physics and Astronomy, 
Manchester, M13 9PL}

\begin{document}

\label{firstpage}

\date{Accepted --; Received --}

\pagerange{\pageref{firstpage}--\pageref{lastpage}}


\maketitle
\begin{abstract}
AMI observations towards CIZA\,J2242+5301, in comparison with observations of weak gravitational lensing and X-ray emission from the literature, are used to investigate the behaviour of non-baryonic dark matter (NBDM) and gas during the merger. Analysis of the Sunyaev-Zel'dovich (SZ) signal indicates
the presence of high pressure gas elongated perpendicularly to the X-ray and weak-lensing morphologies which, given the merger-axis constraints in the literature, implies that high pressure gas is pushed out into a linear structure during core passing. 
Simulations in the literature closely matching the inferred merger scenario show the formation of gas density and temperature structures perpendicular to the merger axis. These SZ observations are challenging for modified gravity theories in which NBDM is not the dominant contributor to galaxy-cluster gravity.

\end{abstract}

\section{Introduction}\label{sec:intro}

Mergers between galaxy clusters can be used to study the behaviour of non-baryonic dark matter (NBDM) and cluster gas (see e.g. \citealt{2001ApJ...561..621R}, \citealt{2004ApJ...604..596C}, and \citealt{2012ApJ...747L..42D}). 
During a merger the NBDM and galaxies interact only gravitationally. The gas components interact through ram pressure and slow as they move closer together. The gas and NBDM components of the merging clusters thus become dissociated from each other. If the merging system is (a) close to head-on between clusters of similar mass, (b) happening approximately in the plane of the sky, and (c) observed shortly after first core passage, then this dissociation may be large enough to be observed. A small number of merging clusters have been reported to show these qualities, such as the Bullet cluster (see \citealt{2004ApJ...604..596C}), A2146 (e.g. \citealt{2016MNRAS.459..517K}) and the Sausage cluster (CIZA\,J2242+5301, references below) with which this paper is concerned.

CIZA\,J2242+5301 ($z$\,=\,0.192; \citealt{2007ApJ...662..224K}) hosts synchrotron-emitting, Mpc-scale radio relics (visible from 0.15\,GHz up to 30\,GHz) that result from shocks from cluster merging (see e.g. \citealt{2010Sci...330..347V}, \citealt{2011MNRAS.418..230V}, \citealt{2013AA...555A.110S}, and \citealt{2016MNRAS.455.2402S}). 

\citet{2010Sci...330..347V}, \citet{2013MNRAS.429.2617O}, \citet{2014MNRAS.440.3416O} and \citet{2015AA...582A..87A} use
X-ray observations to reveal, for example, an elongated X-ray structure, temperature jumps coincident with the radio relics, and the temperature distribution of the cluster.

\citet{2011MNRAS.418..230V} perform hydrodynamical simulations of radio relic formation from mergers and find that CIZA\,J2242+5301 is well described by two approximately equal-mass clusters merging with low impact parameter, with merger axis about north-south and $\leq$10$^\circ$ from the plane of the sky.

Using weak-lensing data, \citet{2015ApJ...802...46J} produce mass distribution maps and estimate a total mass within $r_{200}$ of $\approx$2$\times$10$^{15}$\,\Msun. The mass and X-ray structures have major axes that are orientated with position angles (PA) $\approx$30$^{\circ}$ and $\approx$20$^{\circ}$ respectively. 
Lensing maps show a double-peaked mass distribution, interpreted as the two NBDM components that have passed through each other. 
As expected, the mass distribution peaks coincide with cluster galaxy distribution peaks presented by Jee et al..

\begin{figure*}
\centering 
\includegraphics[trim= 5mm 55.5mm 10mm 51mm, clip, width=80mm]{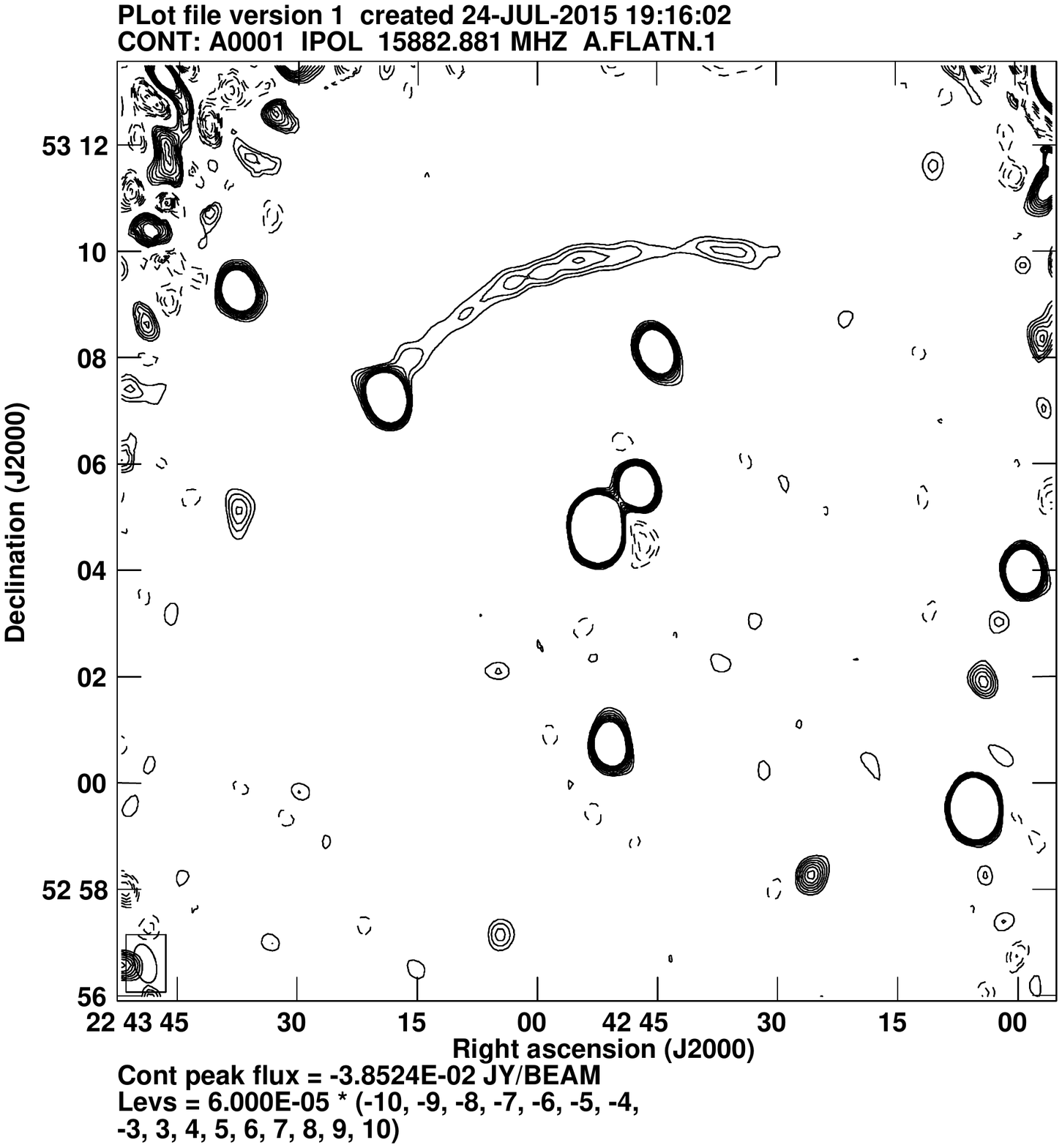}\hspace{0.5cm}
\includegraphics[trim= 5mm 55.5mm 10mm 51mm, clip, width=80mm]{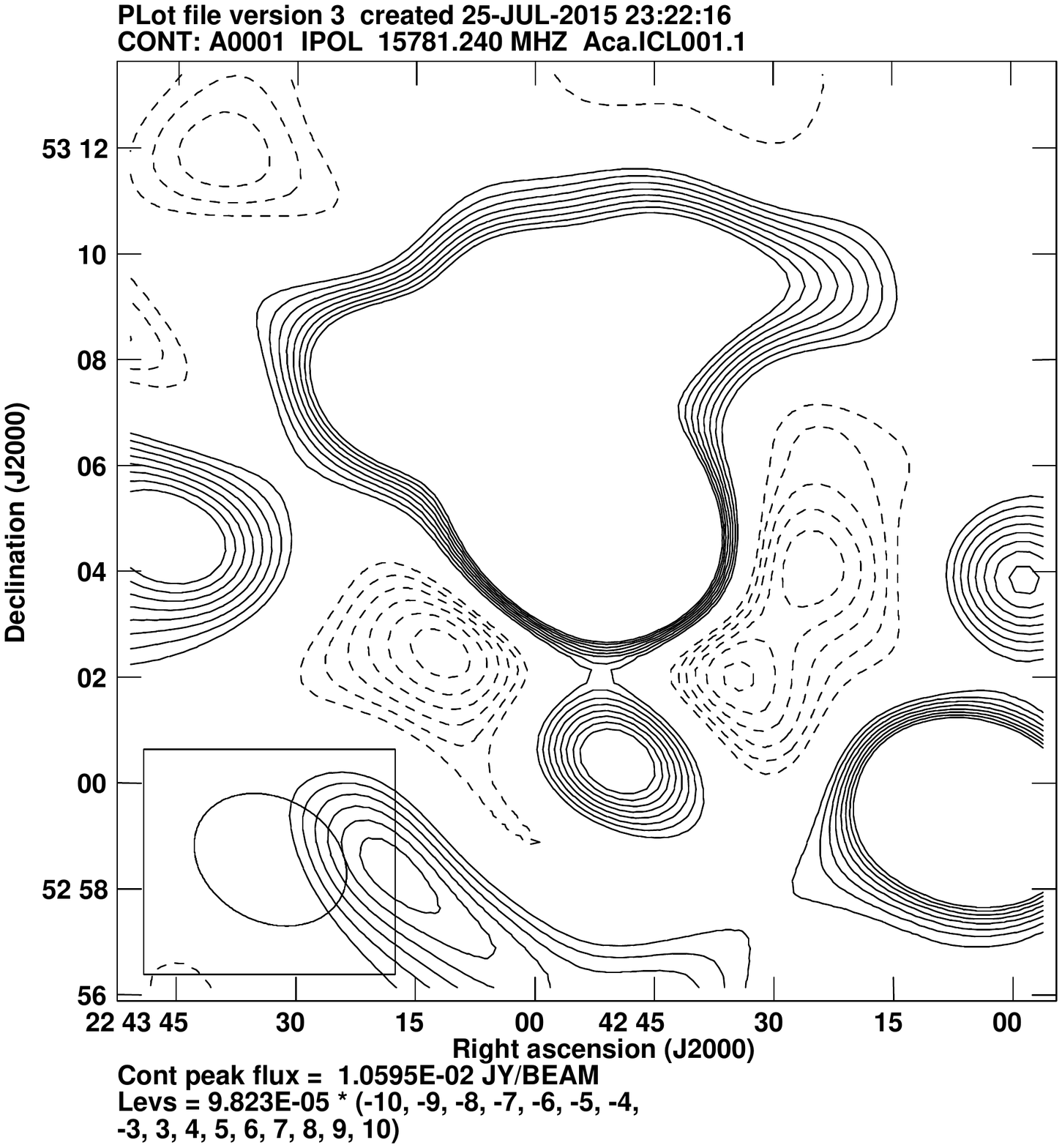}
\caption{Left: Low-noise region of the 61+19 point LA map. Contours are $\pm$3$\sigma_{\rm LA}$ to $\pm$10$\sigma_{\rm LA}$, where $\sigma_{\rm LA}$ is the noise level at the map centre $\approx$60\,$\mu$Jy. At the Sausage relic the noise level is $\approx$25$\mu$Jy. Right: the same region observed by the SA. Contours are $\pm$3$\sigma_{\rm SA}$ to $\pm$10$\sigma_{\rm SA}$, where $\sigma_{\rm SA}$ is the noise at the map centre $\approx$98\,$\mu$Jy. In all AMI maps the half power contour of the synthesised beam is shown in the bottom left-hand corner.}
\label{fig:AMIsausage}
\end{figure*}

\begin{figure}
\centering
\includegraphics[angle=270,trim= 40mm 49mm 25mm 20mm, clip,width=95mm]{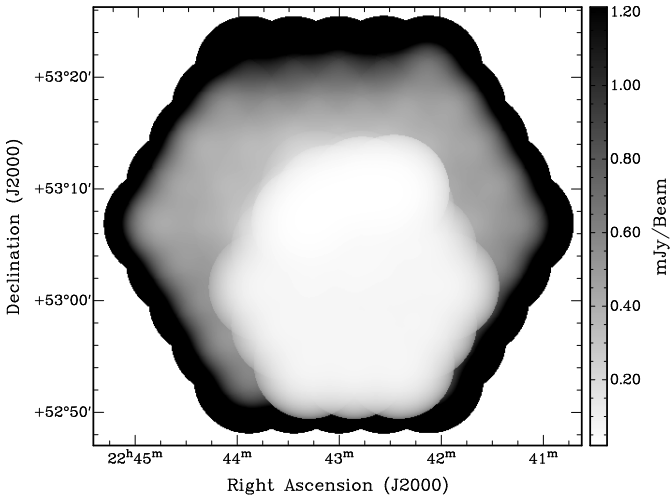}
\caption{Noise map of the LA concatenated data. Noise levels are shown in the colour bar and calculated using the \textsc{aips} task \textsc{imean}. \label{fig:noise}}
\end{figure}

\section{AMI and Observations}
\label{sec:observations}

The Arcminute Microkelvin Imager (AMI) consists of two interferometer arrays: the Small Array (SA), made up of ten 3.7-m antennas with 5--20-m baselines giving 3\textquotesingle~resolution and the Large Array (LA), formed of eight 12.8-m antennas with 18-110\,m baselines giving 30\textquotesingle\textquotesingle~resolution (see e.g. \citealt{2008MNRAS.391.1545Z}).
CIZA\,J2242+5301 was observed with both arrays between July 2012 and February 2013 towards the ROSAT centroid at 13.9-18.2\,GHz with a central frequency of 15.7\,GHz and channel bandwidth of 0.72\,GHz. 
Single-pointed SA observations, covering the positions of both north and south radio relics, were made for a total of 14 hours. Initial LA observations were 61-point mosaics corresponding to the SA primary beam, in which each of the central 19 pointings are observed for 3 times longer than each of the outer pointings. 
Further LA observations were 19-point and 7-point mosaics focusing on the complicated source environment close to the cluster, for a total of 25 hours. A final 8-hour LA observation was made of the Sausage relic itself using 4 points along the length of the structure. This produced the noise distribution shown in Figure \ref{fig:noise}. These observations were used in studies of the radio emission in \citet{2014MNRAS.441L..41S} and \citet{2016MNRAS.455.2402S}.

Flux calibration, data reduction and imaging is carried out as described in \citet{2016MNRAS.460..569R}, summarised here. 
Flux calibration was performed using observations of 3C\,48, 3C\,286 and 3C\,147, with 3C\,286 calibrated against VLA measurements from \citet{2013ApJS..204...19P} and 3C\,48 and 3C\,147 standard flux densities set from long-term AMI monitoring of these sources. Reduction of raw data is carried out in our in-house reduction software \textsc{reduce} where data are flagged for interference, Fourier transformed, calibrated, and written out as $uv$-fits files which are concatenated into a single data set for each array. Imaging of AMI LA and SA data is done using \textsc{aips}. The \textsc{aips} task \textsc{imagr} is used to deconvolve the ``dirty'' map and perform simple \textsc{clean}-ing steps. The task \textsc{imean} is used to determine map noise levels.
These procedures are run in an automated fashion, using natural weighting, to produce an SA map and LA maps for each channel. 
LA and SA maps are shown in Figure \ref{fig:AMIsausage}. 
Mapped LA channel data are used with the \textsc{Source-Find} algorithm \citep{2011MNRAS.415.2699A}, which searches for peak flux densities greater than 4$\sigma_{\rm LA}$ and calculates flux density and spectral index estimates at these positions.

The complementary resolutions of the SA and LA allow both sensitivity to the large-scale SZ signal and subtraction of confusing source environments (see e.g. \citealt{2015AA...580A..95P}, \citealt{2016MNRAS.460..569R}). However, much of the radio emission in the CIZA\,J2242+5301 field is extended so the LA map does not provide an accurate model of the SA source environment. At AMI frequencies the SA map source flux densities are likely to be contaminated by SZ and at low resolution the diffuse material is blended with nearby point sources.
We have therefore used both LA and SA maps to form a model of the SA source environment (Section \ref{sec:sourcemodel}).

%
\section{Data analysis}
\label{sec:analysis}
{\color{black}Analysis} of AMI observations of clusters is usually performed (see e.g. \citet{2016MNRAS.460..569R}) using the Bayesian analysis package \textsc{McAdam} \citep{2009MNRAS.398.2049F} which employs the fast sampler \textsc{multinest} (\citealt{2008MNRAS.384..449F}, \citealt{2009MNRAS.398.1601F} and \citealt{2013arXiv1306.2144F}) to fit simultaneously both a cluster model and a source-environment model to the SA data in $uv$-space. The source-environment model usually consists of delta priors on the source positions and Gaussian priors for the flux densities and spectral indices, typically determined from LA observations.
The cluster model is described in \citet{2012MNRAS.423.1534O} and \citet{2013MNRAS.430.1344O}, and uses NFW-like profiles to describe the gas pressure profile and the dark-matter density profile. The model also assumes a spherical geometry, hydrostatic equilibrium (HSE) within $r_{200}$\footnote{The cluster radii internal to which the mean density is 200 times the critical density at the cluster redshift.} (but not within lower radii), the ``universal'' profile \citep{2010AA...517A..92A}, and requires that the gas fraction is small compared to unity within $r_{200}$ (but not within lower radii).

\begin{table}
\centering
\caption{Point sources in the field of the CIZA\,J2242+5301 system as observed by the LA and identified as point-like by Source-find and visual inspection of the map. Column 1 and 2 show the source J2000 RA and Dec from the LA map, and column 3 shows the LA flux density.}\label{tab:fluxes}
\begin{tabular}{p{2.0cm}p{2.0cm}p{2.1cm}}
\hline
Right Ascension   & Declination                       & Flux density (mJy)             \\
\hline
22~~42~~52.55 &  +53~~04~~47.41 & 10.1 $\pm$ 0.5  \\[1pt]
22~~41~~33.09 &  +53~~11~~06.53 & 8.5 $\pm$ 0.6   \\[1pt]
22~~42~~05.32 &  +52~~59~~31.82 & 6.3 $\pm$ 0.3   \\[1pt]
22~~43~~18.88 &  +53~~07~~15.74 & 2.6 $\pm$ 0.1   \\[1pt]
22~~41~~59.20 &  +53~~03~~58.13 & 2.1 $\pm$ 0.1   \\[1pt]
22~~42~~47.49 &  +53~~05~~35.68 & 2.0 $\pm$ 0.1   \\[1pt]
22~~42~~45.10 &  +53~~08~~07.81 & 1.75 $\pm$ 0.09 \\[1pt]
22~~42~~50.91 &  +53~~00~~44.05 & 1.6 $\pm$ 0.1   \\[1pt]
22~~43~~37.58 &  +53~~09~~16.11 & 1.4 $\pm$ 0.1   \\ [1pt]
22~~42~~25.77 &  +52~~58~~16.80 & 0.58 $\pm$ 0.08 \\[1pt]
22~~42~~04.18 &  +53~~01~~53.55 & 0.46 $\pm$ 0.07 \\[1pt]
22~~43~~37.38 &  +53~~05~~07.91 & 0.38 $\pm$ 0.05 \\ [1pt]
22~~43~~04.65 &  +52~~57~~09.13 & 0.33 $\pm$ 0.07 \\[1pt]
22~~43~~04.82 &  +53~~02~~06.71 & 0.25 $\pm$ 0.06 \\[1pt]
22~~43~~03.04 &  +53~~07~~47.38 & 0.17 $\pm$ 0.03 \\[1pt]
22~~43~~16.84 &  +53~~05~~51.74 & 0.15 $\pm$ 0.03 \\[1pt]
\hline
\end{tabular}
\vspace{0.3cm}
\end{table}

\begin{table*}
\centering
\caption{Summary of the priors used on source parameters for point source and sources modelling diffuse emission. These priors are used in all analyses detailed in Table \ref{cluster_priors}.}\label{source_priors}
\begin{tabular}{p{4.5cm}p{12cm}}
\hline
Point source parameter                          & Prior                 \\
\hline
Position $(x_{\rm s},y_{\rm s})$        & Delta-function at LA position                                                         \\[2pt]
Flux density $(S_0/Jy)$ $>$\,4$\sigma_{\rm SA}$    & Gaussian at LA value, with $\sigma$\,=\,40 per cent of LA flux density      \\[2pt]
Spectral index when $S_{0}$ $>$\,4$\sigma_{\rm SA}$  & Gaussian centred at LA-fitted spectral index with $\sigma$\,=\,LA error, or prior based on 10C; see Section \ref{sec:analysis} \\[2pt]
Flux density $(S_0/Jy)$ $<$\,4$\sigma_{\rm SA}$    & Delta-function at LA value, unless close to cluster            \\[2pt]
Spectral index when $S_{0}$ $<$\,4$\sigma_{\rm SA}$  & Delta-function centred at LA-fitted spectral index, or based on 10C; see Section \ref{sec:analysis}         \\
\hline
Extended emission parameter                          &                 \\
\hline
Position $(x_{\rm s},y_{\rm s})$  & Delta-function at position based on LA and SA map position        \\[2pt]
Flux density $(S_0/Jy)$           & Gaussian at SA value, with $\sigma$\,=\,40 per cent of SA flux density      \\[2pt]
Spectral index                    & Gaussian centred at spectral index based on 10C or \cite{2013AA...555A.110S} and \cite{2016MNRAS.455.2402S} \\
\hline
\end{tabular}
\vspace{0.3cm}
\end{table*}

\begin{table*}
\centering
\caption{Summary of the priors used on cluster parameters in the model described in Section \ref{sec:analysis}. $M_{{\rm tot},\,r_{200}}$ and $f_{{\rm gas},\,r_{200}}$ are the total mass internal to $r_{200}$ and gas fraction internal to $r_{200}$, respectively. Three analyses are run with the full source environment. Analyses done to form the full source environment model use the priors in column 2 (see Section \ref{sec:sourcemodel}).}\label{cluster_priors}
\begin{tabular}{p{2.0cm}p{4.5cm}p{4.5cm}p{4.5cm}}
\hline
Parameter                          & Prior - set I   & Prior - set II &    Prior - set III               \\
\hline
Cluster~position $(x_{\rm c},y_{\rm c}/ \rm{arcsec})$       & Single Gaussian at SA pointing centre, $\sigma$\,=\,180 arcsec  & Double Gaussian at lensing peaks, $\sigma$\,=\,60 arcsec  &Double Gaussian at SZ mode positions, $\sigma$\,=\,60 arcsec  \\[5pt]

Mass $(M_{{\rm tot},\,r_{200}}$~/\Msun$)$        & Uniform in log space between 1$\times10^{14}$ and 6$\times10^{15}$  & Uniform in log space between 1$\times10^{14}$ and 6$\times10^{15}$ for each cluster  & Uniform in log space between 1$\times10^{14}$ and 6$\times10^{15}$ for each cluster       \\[5pt]

Gas fraction $(f_{{\rm gas},\,r_{200}})$        & Gaussian with $\mu$\,=\,0.13, $\sigma$\,=\, 0.02          &  Gaussian for each cluster with $\mu$\,=\,0.13, $\sigma$\,=\, 0.02 &  Gaussian for each cluster with $\mu$\,=\,0.13, $\sigma$\,=\, 0.02           \\[5pt]

Shape parameters $(a, b, c, c_{500})$          & Delta-functions with ``universal'' values, see \citet{2010AA...517A..92A}     &  Delta-functions with ``universal'' values  & Delta-function with ``universal'' values      \\
\hline
\end{tabular}
\vspace{0.3cm}
\end{table*}

Given previous observations of CIZA\,J2242+5301 (see Section \ref{sec:intro}), available analytical cluster models, such as that described above and in \citet{2016MNRAS.460..569R}, are likely a poor fit to the SZ signal. We therefore adapt the usual analysis to deal more accurately with the diffuse emission in the field and the very disturbed, merging system. This is achieved by adopting a three-stage process. The first two stages consist of constructing a model for the point sources and diffuse emission to obtain posterior estimates of the source environment from LA and SA data, providing delta priors on source positions and Gaussian priors on flux densities for the analyses in Section \ref{sec:cluster}.
The final stage uses this ``source-environment'' model in further analyses to search for possible multiple SZ centres associated, for example, with the two NBDM concentrations, to provide a potentially improved fit to the observations (see Section \ref{sec:cluster}).

\subsection{Source-environment modelling}\label{sec:sourcemodel}
{\color{black}With} a resolution of $\approx$3\,arcmin, the SA map shows much of the large-scale radio emission in the field, notably the Sausage relic with a peak surface brightness of around 3\,mJy\,beam$^{-1}$. The point sources in the field are blended with the large-scale radio emission.
With a resolution of $\approx$30\,arcsec, the LA map resolves the Sausage relic from surrounding sources and resolves out the SZ effect and much of the larger scale radio emission, giving a much lower peak brightness of $\approx$350\,$\mu$Jy\,beam$^{-1}$. We construct a source-environment model that describes both point sources and large-scale radio emission using both the LA and SA data, with an iterative two-part \text{McAdam} analysis, as outlined below.

\subsubsection{Point-source-only modelling}\label{sec:pointonly}
First, map-plane \textsc{Source-Find}-ing of LA data identifies local maxima with flux values above 4$\sigma$; this includes not only point sources in the LA map but also resolved peaks of the extended emission in the Sausage relic and to the east and south of the merger. Thus, only sources visually determined to be point sources from the LA map and higher resolution maps in e.g. \citet{2013AA...555A.110S} are retained. The positions of these sources and the LA flux density estimates are detailed in Table \ref{tab:fluxes}.

This point-source model is used as input to \textsc{McAdam}. The priors on the point source parameters are summarised in Table \ref{source_priors}. 
Point-source position priors are delta functions on the LA position, and Gaussian priors are placed on the LA flux estimate and the spectral index (either from the LA channel data or based on the 10C survey). Faint point sources far from the expected SZ signal and the diffuse emission are given delta priors on both position and flux density.

The priors on cluster parameters used in \textsc{McAdam} for this analysis are given in Table \ref{cluster_priors} (set I). 
The prior on cluster mass internal to $r_{200}$, $M_{{\rm tot},\,r_{200}}$, is uniform in log in the range $1\times10^{14}$\Msun~-- $6\times10^{15}$\Msun, despite previous mass estimates from \citet{2011MNRAS.418..230V} and \citet{2015ApJ...802...46J}, to allow parameter estimates to be as data driven as possible.
Given the large angular extent of the merger shown in X-ray and lensing maps it is not clear where in the system the peak SZ will be seen, so in initial analyses a Gaussian cluster position prior with $\sigma$\,=\,180\textquotesingle\textquotesingle~is used on the map centre (Table \ref{cluster_priors}, set I).
The prior on the gas fraction internal to $r_{200}$, $f_{{\rm gas},\,r_{200}}$, is a narrow Gaussian centred at 0.13 with $\sigma$=0.02 (see e.g. \citealt{2006ApJ...640..691V}, \citealt{2009ApJ...692.1033V} and \citealt{2011ApJS..192...18K}) and the pressure-profile shape parameters are given delta priors at the ``universal'' values of \citet{2010AA...517A..92A}.

The point-source parameters produced from this analysis are then subtracted from the SA $uv$-data resulting in the map shown in the left of Figure \ref{fig:saus}. Comparing with the right panel in Figure \ref{fig:AMIsausage} the diffuse emission appears no longer blended with the point sources and is seen in three regions of the map: the most significant is the Sausage relic itself to the north of the map; another region to the east of the cluster appears to coincide with extended emission in lower frequency maps (see e.g. \citealt{2013AA...555A.110S}); a small amount of emission is seen to remain from the ``Southern Sausage'' visible in lower frequency maps to the south and south-west of the cluster. There is also an area of unexplained emission to the south-east, noted by \citet{2014MNRAS.441L..41S}, which does not appear to correspond to lower frequency emission.

\subsubsection{Full source-environment modelling}\label{sec:fullsource}
The diffuse emission remaining in the SA map is then modelled as collections of point sources separated by between 1.2 and 1.6 arcmin -- half the dimensions of the minor and major axes of the half-power contour of the synthesised beam (shown in the bottom left of the SA maps in Figure \ref{fig:AMIsausage} and \ref{fig:saus}). The positions of these ``mock'' sources are chosen by visual inspection of the diffuse emission. 
Estimates of the flux density of the ``mock'' sources are made from the point-source-subtracted SA map (Figure \ref{fig:saus}) at these positions and used as priors. 
A simulated SA observation of the seven sources used to model the Sausage relic verifies that they are equivalent to a line source at the resolution of the SA, approximating the Sausage relic. Given the SA synthesised beam, shown in Figure \ref{fig:beam}, sidelobes from the large decrement may be boosting the diffuse emission by up to 10 per cent.

The ``mock'' sources approximating the extended emission are added to the point-source list with wide priors to allow for the possible SZ contamination of SA-map flux densities and for the possible contribution of sidelobes from the SZ signal. 

This combined source-environment model is then used as input to a second \textsc{McAdam} analysis. The priors assumed on the parameters of these sources are summarised in Table \ref{source_priors}. The priors on the cluster parameters are the same as those used in the first \textsc{McAdam} analysis (Table \ref{cluster_priors}, set I).

After subtracting the flux-density estimates obtained in this second \textsc{McAdam} analysis, almost all of the extended emission appears to have been accurately removed, as shown in the right-hand panel of Figure \ref{fig:saus} (which also shows the positions of the full source list).

\subsection{SZ measurement}\label{sec:cluster}

In all SA maps of CIZA\,J2242+5301, the SZ decrement is clearly non-circular and always appears elongated east-west. Several test analyses have been conducted using the same cluster and point-source priors as previously but with different modelling for the areas of extended emission. The cluster-parameter estimates do depend on the diffuse-emission modelling, but this dependence cannot account for the double-structure and elongation of the SZ decrement.

All analyses find two modes i.e. two areas of the parameter space with high probability values; the Bayesian evidence values of each mode in a pair are always close and thus do not suggest one mode is dominant. Posterior probability distributions for the analysis in which the full source-environment modelling is used are shown in Figure \ref{fig:tri}. The distributions in position are clearly bi-modal; probability distributions of other parameters are similar between the modes which are marginalised in Figure \ref{fig:tri}. We will refer to the mean position estimates of the two modes from this analysis as the ``SZ mode positions'' from here on.

\begin{figure*}
\centering 
\includegraphics[trim=5mm 55.5mm 10mm 51mm, clip, width=80mm]{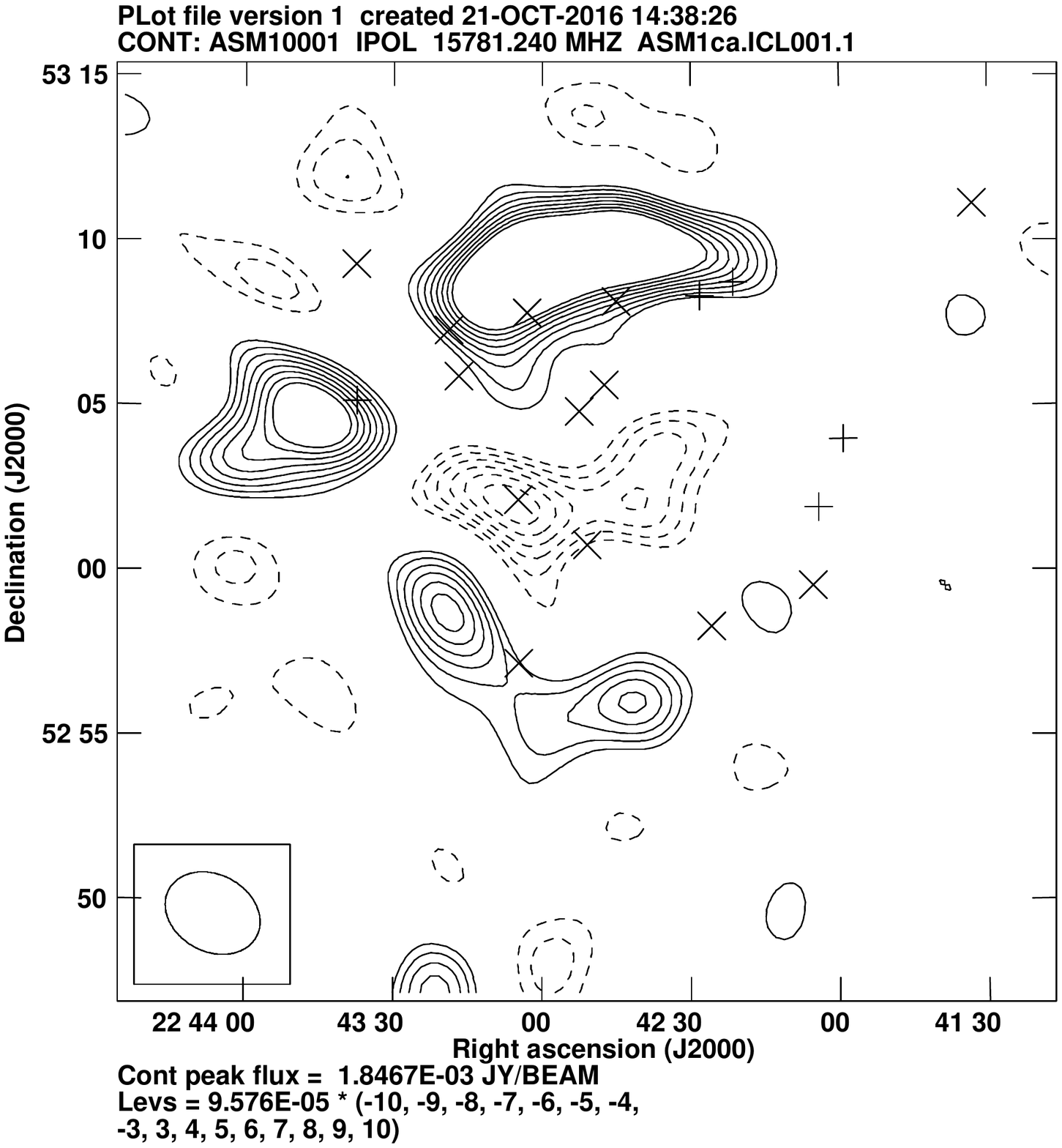}\hspace{0.5cm}
\includegraphics[trim=5mm 55.5mm 10mm 51mm, clip, width=80mm]{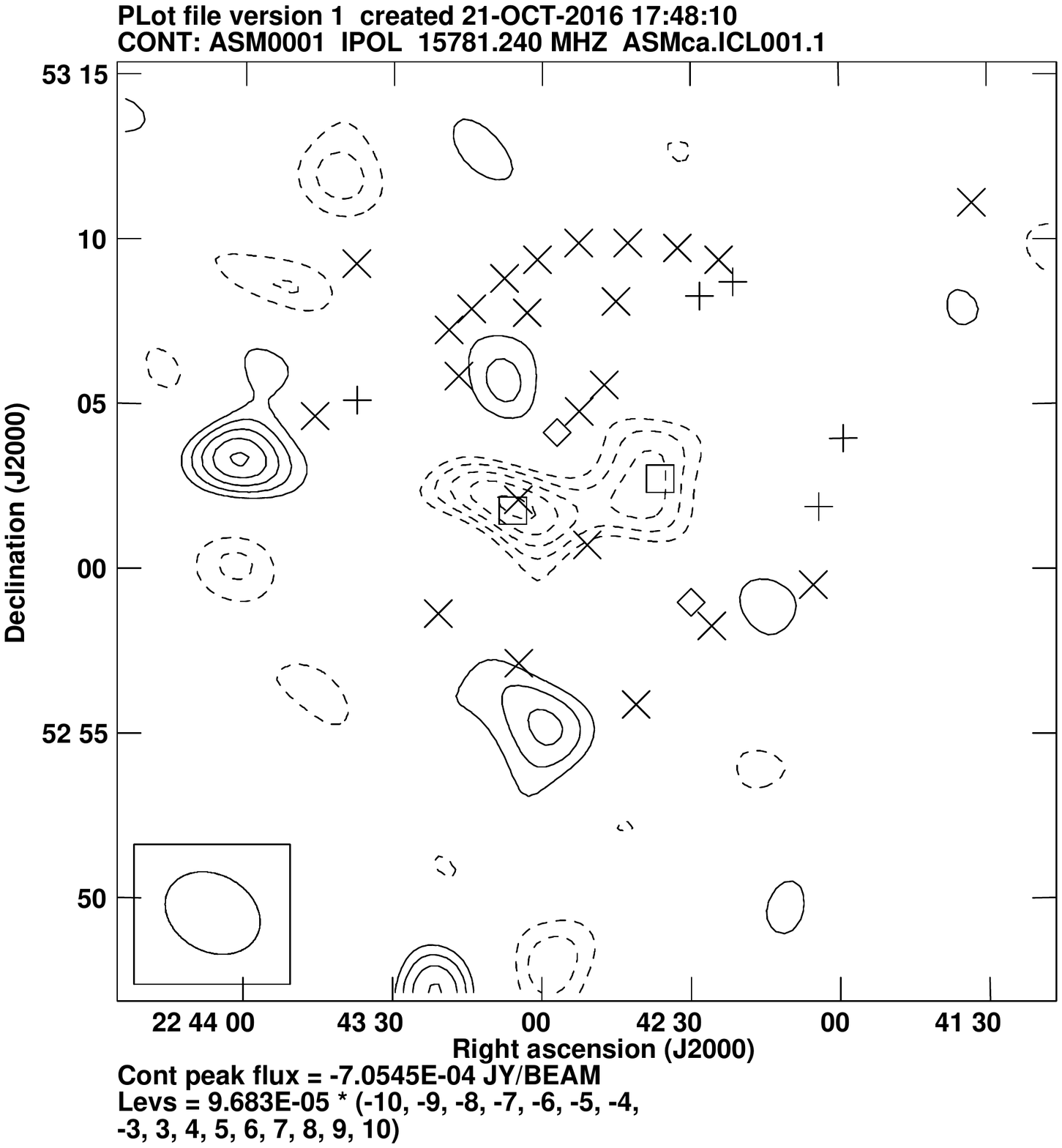}
\caption{Left: SA map of CIZA\,J2242+5301 after subtraction of LA point sources ($\sigma$\,$\approx$\,96\,$\mu$Jy). Right: full source-environment modelled in the analysis and subtracted ($\sigma$\,$\approx$\,97\,$\mu$Jy). Contours are $\pm$3$\sigma$ to $\pm$10$\sigma$. `$\times$' show sources that are given Gaussian priors on flux density and spectral index, and `$+$' show those modeled with delta priors. Squares show fitted cluster positions, diamonds show lensing-peak positions.}
\label{fig:saus}
\end{figure*}

\begin{figure}
\centering 
\includegraphics[trim= 15mm 50mm 10mm 43mm, clip, width=78mm]{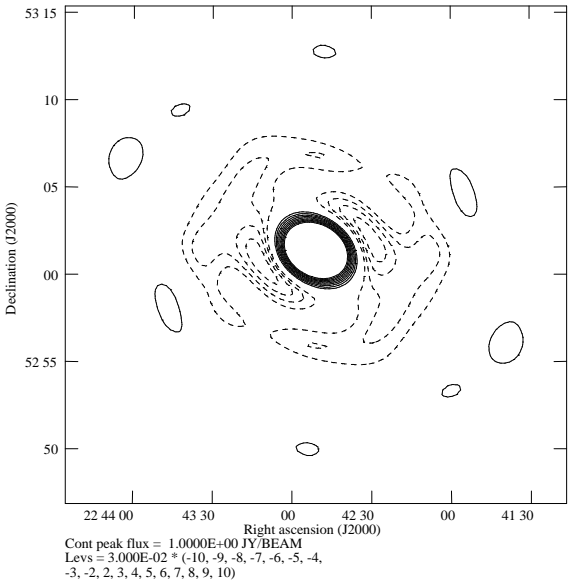}
\caption{Synthesised beam shape of the SA for CIZA\,2242+5301 observations detailed in Section \ref{sec:observations}.}
\label{fig:beam}
\end{figure}

\begin{figure}
\centering
\includegraphics[trim= 10mm 5mm 20mm 88mm, clip,width=90mm]{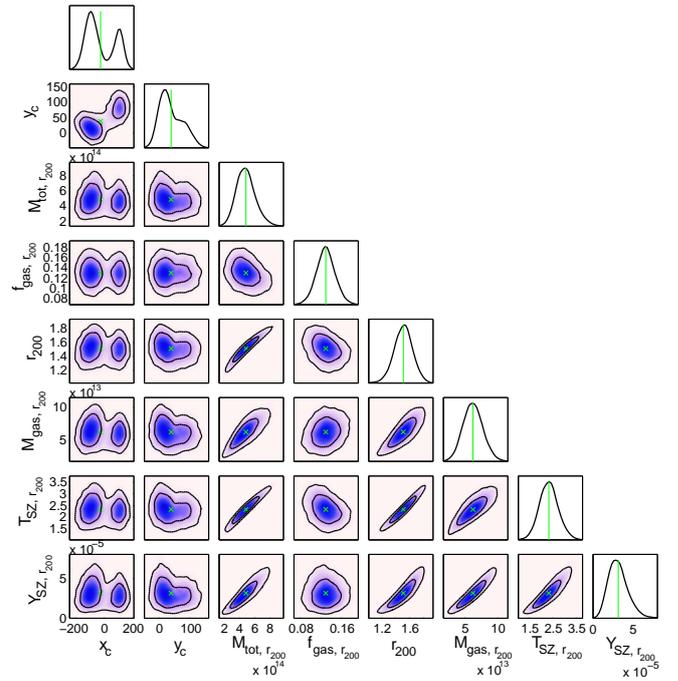}
\caption{Marginalised posterior probability distributions for some key fitted parameters for the bi-modal analysis, using the full source environment model. $M_{{\rm tot},\,r_{200}}$ and $f_{{\rm gas},\,r_{200}}$ are the total cluster mass within $r_{200}$ and the gas fraction within $r_{200}$, as shown in Table \ref{cluster_priors}. $M_{{\rm gas},\,r_{200}}$ is the gas mass within $r_{200}$, $T_{\rm SZ}(r_{200})$ (in keV) is the gas temperature at $r_{200}$, and $Y_{SZ,\,r_{200}}$ (in ${\rm Mpc}^{-2}$) is the Comptonization parameter integrated over the angle subtended by the cluster. $r_{200}$ is given in Mpc. \label{fig:tri}}
\end{figure}

We have investigated if there are gas mass peaks coincident with the lensing-mass peaks, hidden in the previous analysis by the source environment. Accordingly, we model two spherical clusters instead of one by using two sets of cluster priors with position priors centred on the two \citet{2015ApJ...802...46J} lensing peaks each with $\sigma$= 60\textquotesingle\textquotesingle\,(Table \ref{cluster_priors}, set II).
The parameter values fitted are similar to those of the modes but the position posteriors are pulled away from the priors (diamonds in Figure \ref{fig:saus}) towards the SZ mode positions (squares in Figure \ref{fig:saus}). Similarly, the source-subtracted map shows the same elongated decrement as the previous analyses.
When repeated using the SZ mode positions as position priors (Table \ref{cluster_priors}, set III), parameter estimates and SA map decrement are very little different compared to the bi-modal fit (probability distributions are given in Figure \ref{fig:tri} and SA map decrement is shown in the right panel of Figure \ref{fig:saus}). 
Whether modelled as single-structured (using priors for one cluster at the pointing centre) or double-structured at two different PAs (using two sets of cluster priors either at PA$\approx$30$^\circ$ or $\approx$100$^\circ$), the cluster posteriors are always very similar, giving total mass estimates of about 1$\times$10$^{15}$\,\Msun, and resulting in an SZ decrement close to that seen in Figure \ref{fig:saus}.

\begin{figure}
\centering
\includegraphics[trim= 10mm 0mm 20mm 88mm, clip,width=89mm]{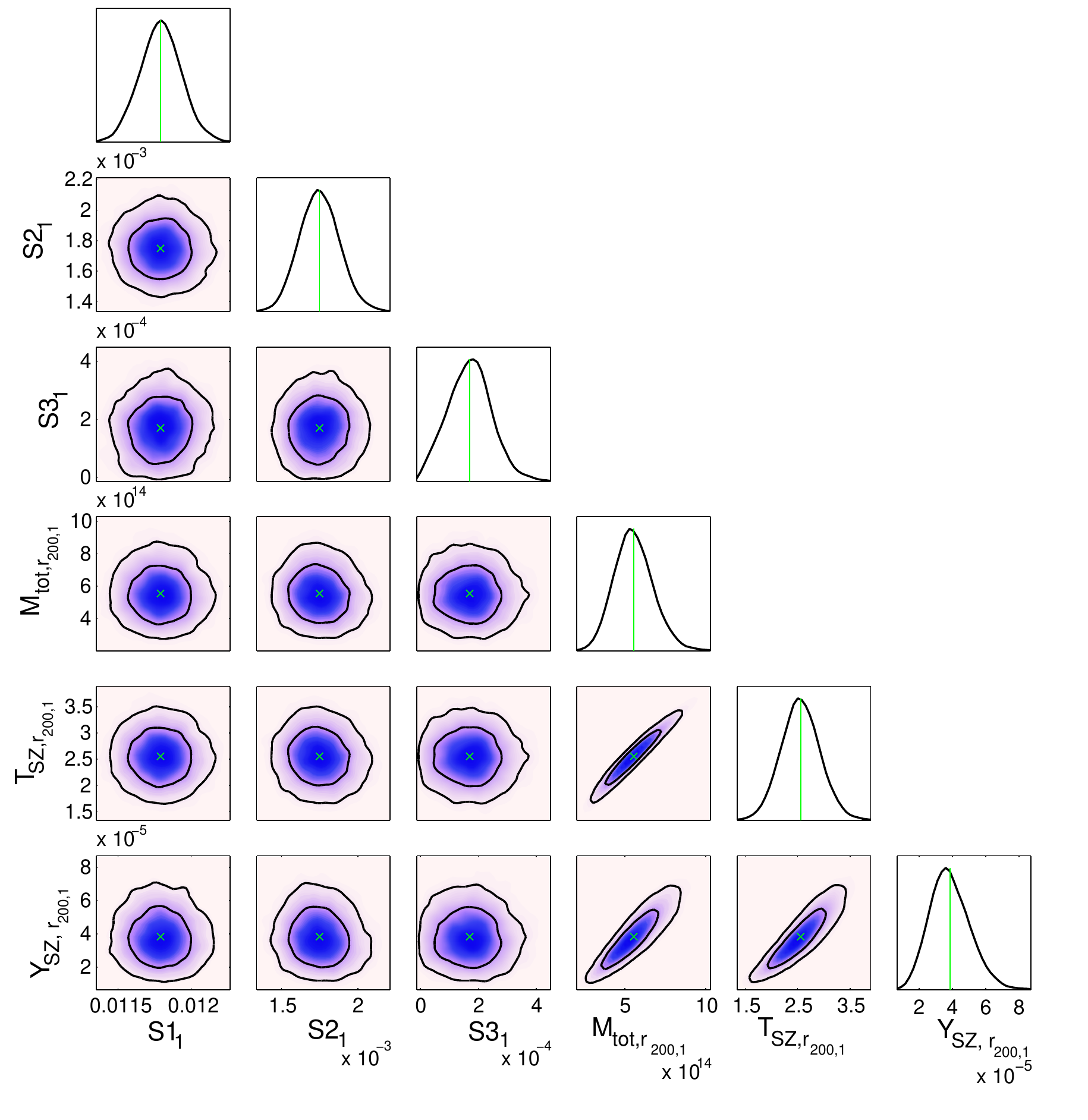}
\includegraphics[trim= 10mm 0mm 20mm 88mm, clip,width=89mm]{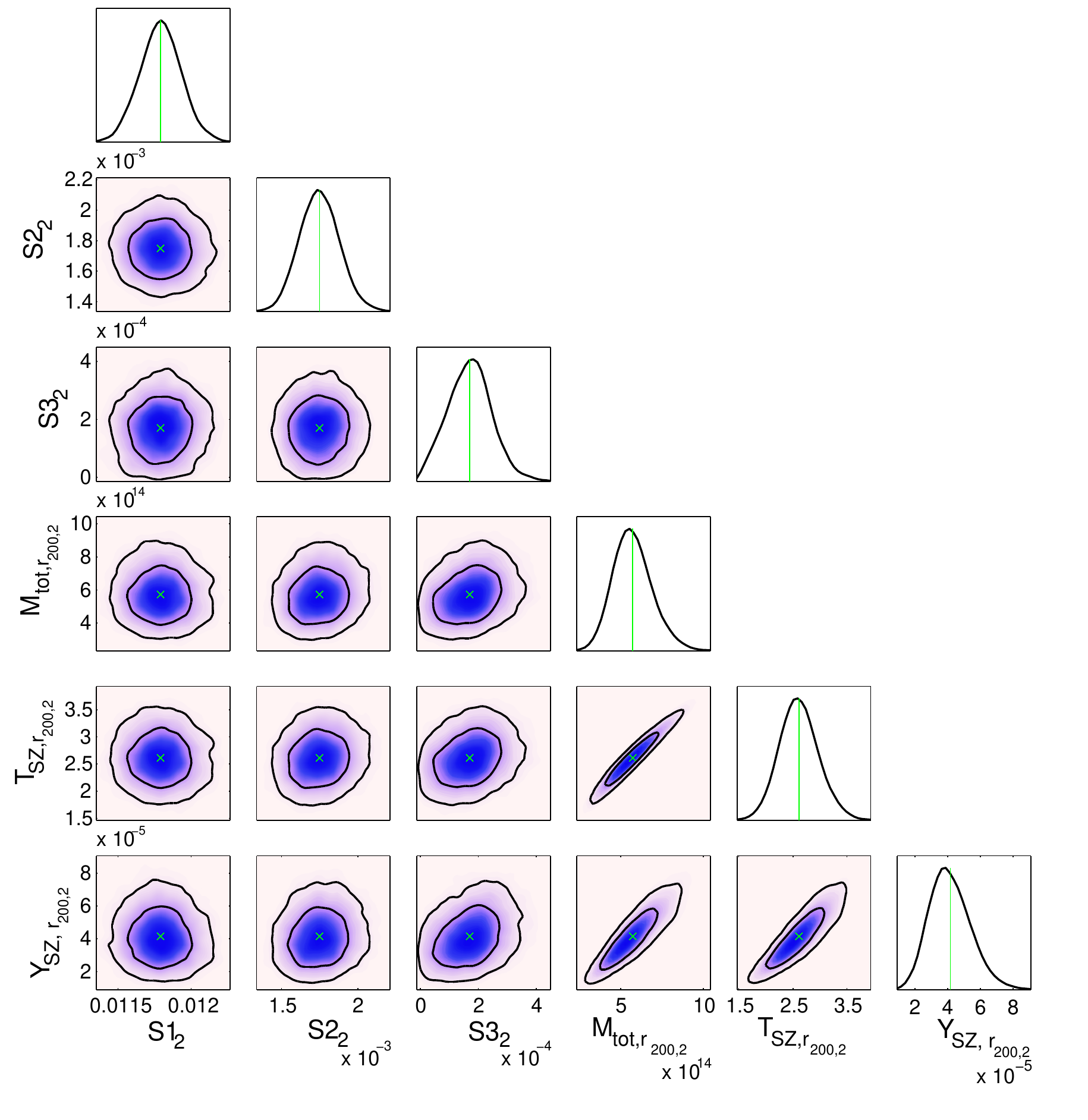}
\vspace{-0.5cm}
\caption{Posterior probability distributions for parameter values of the two clusters modelled at the SZ mode positions (denoted 1 and 2 in the subscripts), using a full source environment model, and the flux densities, in mJy, of the three sources closest to the decrement (S1, S2 and S3 in order of decreasing flux density). \label{fig:tri_src}}
\end{figure}

Analysis of the SZ decrement therefore strongly indicates a double-structured signal that is elongated along PA$\approx$100$^{\circ}$ with a length of some 1.7\,Mpc. For all analyses, a very small amount of degeneracy is seen between cluster parameter estimates and the flux densities of the three closest point sources to the decrement (sources shown by the crosses in Figure \ref{fig:saus}; LA flux densities given in rows 2, 8 and 14 of Table \ref{tab:fluxes}). These degeneracies are shown in the marginalised probability distributions for the two-cluster analysis using the SZ mode positions (Table \ref{cluster_priors}, set III), in Figure \ref{fig:tri_src}, labelled S1, S2 and S3 in order of decreasing flux density. The low degeneracy confirms that our cluster analysis is not significantly affected by the presence of these sources, due to the large angular extent of the system and the relatively low flux density of the sources.

The analyses conducted have tested how the diffuse emission and different cluster structures in the modelling impact on the resulting parameter estimates and map decrement: these show that the SZ shape has not been produced or biased by the source environment or the modelling. However, it is reasonable to expect a small amount of bias in the decrement depth -- see Figure \ref{fig:beam} -- and consequent bias in parameter estimates. Nevertheless, the bias introduced into the parameter estimates by the source environment is small compared with other sources of bias described in rest of this paragraph.
The SZ structure is not consistent with the ``universal'' profile and the assumed geometry and HSE (detailed in Section \ref{sec:analysis}). 
Although \citet{2016MNRAS.460..569R} show that our analysis pipeline is robust against some departures of clusters from the ``universal'' profile, the merger stage here cannot be accommodated.
In addition, the angular size of the cluster is problematic for AMI. SZ signal missed due to baseline length depends on $uv$-coverage and true-surface brightness distribution; Gaussian circular emission with FWHM 3\textquotesingle~will be negligibly attenuated by the SA while such emission with FWHM 6.5\textquotesingle~will be attenuated by a factor $\approx$ 2. However, emission 6.5\textquotesingle~long but narrow -- similar to the SZ seen -- will suffer little attenuation.
The SA maps are thus indeed able to show a narrow structure of high line-integrated-pressure gas in the ICM. 
It is reasonable to expect the SA mass estimates to be biased low, and assume the mass of the whole merging system given by \citet{2015ApJ...802...46J} from weak lensing as $\approx$2$\times$10$^{15}$\,\Msun~to be the most accurate so far. 
 
\section{Comparison with lensing, X-ray, and radio relics}\label{sec:lensingxray}

\begin{figure*}
\centering
\centerline{\includegraphics[trim= 0mm 0mm 0mm 0mm, clip, width=125mm]{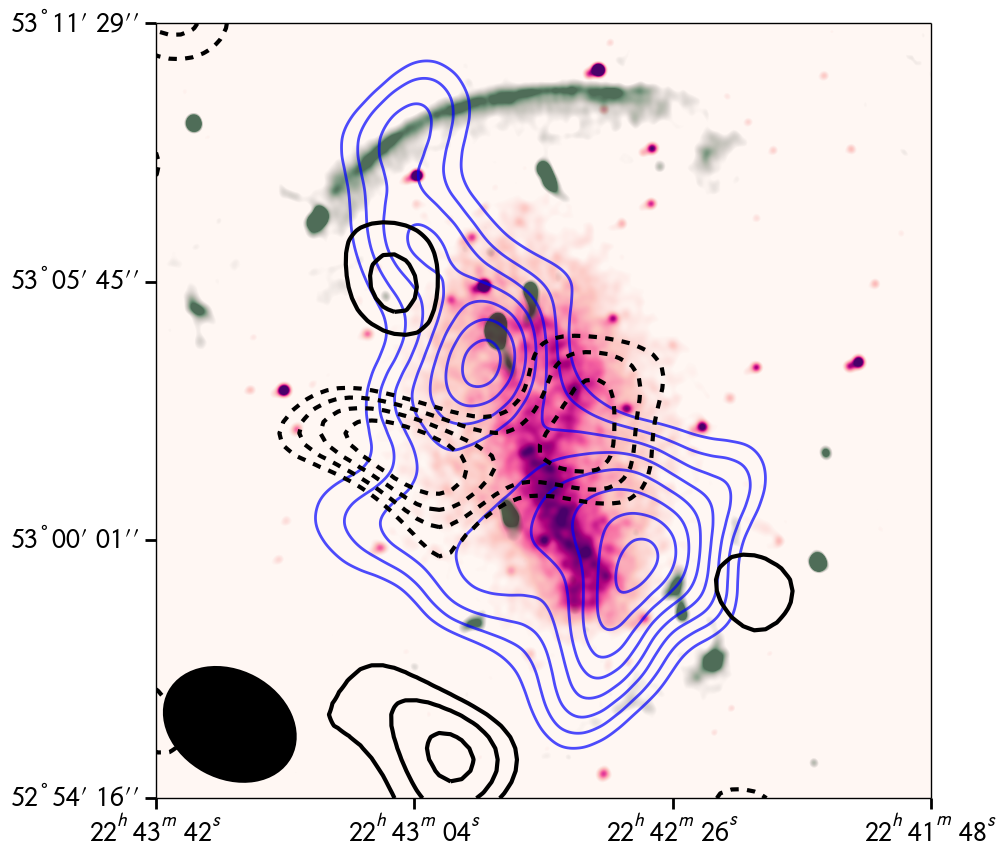}}
\caption{Pink shows $Chandra$ data from \citep{2014MNRAS.440.3416O}. Green shows 300\,MHz GMRT radio emission \citep{2013AA...555A.110S}. Blue contours show the weak-lensing mass distribution of \citep{2015ApJ...802...46J}. Black contours are at $\pm$3$\sigma_{\rm SA}$ to $\pm$6$\sigma_{\rm SA}$, where dashed denotes negative.}
\label{fig:alldata}
\end{figure*}

The X-ray and lensing PAs are consistent with that inferred from the radio relics; these are close to perpendicular to the PA of the SZ decrement.
Figure \ref{fig:alldata} shows a composite of Chandra X-ray \citep{2014MNRAS.440.3416O}, weak lensing \citep{2015ApJ...802...46J}, radio emission, and AMI SZ.

\subsection{Estimating the merger stage}\label{sec:mergerstage}
The merger history of CIZA\,J2242+5301 is discussed in e.g. \citet{2011MNRAS.418..230V}, \citet{2014MNRAS.445.1213S}, \citet{2015ApJ...802...46J} and \citet{2015AA...582A..87A}. 
van Weeren et al. use simulations of binary mergers to replicate shock and relic formation and find agreement of simulated shocks with the CIZA\,J2242+5301 relics using a 2:1 merger with a $\lessapprox$\,400-kpc impact parameter. Using a total mass based on X-ray luminosity and the cluster approach speed calculation of \citet{2001ApJ...561..621R}, van Weeren et al. estimate shock Mach numbers and, from the relic positions, find a merger stage of 1.0\,Gyr since first core passing. Given the expected extinction from the high H\textsc{i} column density \citep{2007ApJ...662..224K} affecting the X-ray-based mass estimate, and the mass estimate of Jee et al. (around 4$\times$ higher), the merger stage is likely to be earlier.
\citet{2014MNRAS.445.1213S} use radio estimates of the electron-injection index to find a shock speed of $\approx$2500\,km\,s$^{-1}$, and estimate that the merger is 0.6-0.8\,Gyr after first core passing. Akamatsu et al. use X-ray measurements of the ICM temperature across the shocks to calculate shock speeds, and estimate the system is $\approx$0.6\,Gyr after first core passing.

We have used the same method as \citet{2011MNRAS.418..230V} and \citet{2015AA...582A..87A}, along with the \citet{2015ApJ...802...46J} mass estimates, to find a relative cluster approach velocity of $v\approx$\,2400\,km\,s$^{-1}$. The two peaks in the lensing-mass reconstructions and galaxy-number density maps of Jee et al. are separated by $\approx$1.4\,Mpc; assuming the point of core passing is half of this separation and that the merging clusters have the same speed, the time since core passing is $\approx$0.57\,Gyr.

\subsection{Gas and dark-matter behaviour during cluster merging}\label{sec:gasdarkmatter}

Figure \ref{fig:alldata} shows a composite map of the X-ray (pink), weak-lensing (blue contours), SZ signal (black contours -- dashed is negative), and 300-MHz emission (green). The X-ray surface brightness appears to have two regions, the southern region elongated approximately along the merger axis and the northern region more diffuse. Given the relative positions of the lensing peaks, tracing the NBDM centres of mass, and the expected merger scenario (Sec. \ref{sec:mergerstage}) of $\approx$0.6\,Gyr since core passing, we interpret these X-ray regions as the progenitor gas cores that have passed each other and dissociated from their NBDM components, which have moved further along their orbital paths.

We suggest the following explanation for the SZ morphology, which lies between the lensing peaks, orientated close-to perpendicular to the lensing and X-ray PAs. Before core passing, the gas pressure between the clusters increases in a disc-like structure as gas slows and builds up due to ram pressure and shocking. Gas is constrained to move down the pressure gradient, perpendicularly away from the merger axis.
During core passing, the progenitor gas cores orbit past each other close to the centre of the disc, displacing disc gas.
Displacement of gas could re-shape the disc-like region into a structure similar to a torus. The two SZ peaks thus may correspond to the two parts of the torus that have the highest line-of-sight integrals of gas pressure.
The density and temperature are likely to be higher towards the progenitor gas cores, but we detect no SZ signal towards the southern X-ray peak. The proximity of the western SZ peak to the northern X-ray peak could indicate that there is some contribution to the SZ emission from this gas. 
Due to the lower dependency of the SZ signal to the gas number density compared to X-ray emission, and the smaller solid angles subtended by the gas cores compared to the area of the synthesised beam, they are not visible to the SA. As evidenced in Figure \ref{fig:alldata}, SZ from the southern X-ray peak will suffer greater beam dilution than the northern peak, so the SZ decrement visible to the SA is dominated by the gas flow perpendicular to the merger axis.

The observations agree well with \citet{2001ApJ...561..621R}, who simulate NBDM and gas surface densities and X-ray surface brightness for equal mass mergers with a small impact parameter. The simulated gas surface density and temperature distributions show structure, perpendicular to the merger axis, forming during core passing (at an angle consistent with the PA of the observed SZ), whereas the X-ray surface brightness is dominated by the gas cores and does not show perpendicular features (see the centre panel of Figures 3 and 4 of Ricker \& Sarazin). Similar perpendicular structures are also seen in simulations by \citet{2006MNRAS.373..881P} and \citet{2011MNRAS.418..230V}. This is also in agreement with recent simulations by \citet{2017arXiv170305682D}, who simulate CIZA\,J2242+5301 in order to determine masses of the component clusters that would form a model consistent with the majority of observations.

This SZ observation appears challenging for gravity theories with no NBDM on galactic and greater scales, for example with the gravitational constant G that is $\approx$10$\times$ (at the present epoch) larger than in the Solar neighbourhood, (see e.g. \citealt{1983ApJ...270..365M} and \citealt{2016arXiv160609128I}). In such a scenario, the gas is gravitationally dominant; this is not the case in CIZA\,J2242+5301 given the following. First, the weak lensing will be dominated by the gas and, since SZ surface brightness (for isothermal gas) traces gas surface density, the weak-lensing-deduced mass and SZ distributions should coincide. Second, there would be a far greater gravitational opposition to the lateral gas expansion so SZ, weak-lensing and X-ray distributions would be more circular and co-centred than is observed. Given the well studied merger scenario and the orientation of the merger axis with the line of sight, this is a clearer gas and dark-matter dissociation than in other mergers such as e.g. the Bullet, where the recognisable X-ray shock feature and the western NBDM are nearly in the same direction.

\section{Conclusions}

\begin{enumerate}

   \item SA observations of major merger CIZA\,J2242+5301 are analysed in a Bayesian way using a multiple NFW-like cluster model and a source environment model based on LA and SA maps. The posterior source-environment model is subtracted to show an SZ signal from CIZA\,J2242+5301 that is elongated along PA$\approx$100$^{\circ}$, close to perpendicular to the merger axis revealed from X-ray, radio relics, and weak-lensing maps.

   \item Given the \citet{2011MNRAS.418..230V} merger scenario and the weak-lensing mass distribution peak positions, we have used the \citet{2015ApJ...802...46J}. mass estimate to calculate a time since first core passing of $\approx$0.6\,Gyr.

   \item The lensing peaks, showing the non-baryonic dark matter (NBDM), are closely followed by the X-ray emission from the progenitor gas cores. The SZ signal visible to the SA shows a high pressure region of the cluster gas and appears as a double peaked decrement orientated close-to perpendicularly to the axis of the lensing and X-ray signals, and positioned between the lensing mass peaks.
 Given the estimated merger stage and scenario, the SZ decrement morphology implies the following: a disc-like region of high pressure gas, formed as the progenitors merged, may have become torus-like when the progenitor gas cores orbited past each other near the centre of the disc, displacing the disc-gas away from the disc centre. This is consistent with high density and temperature structures seen perpendicular to the merger axis in simulations of equal-mass mergers with small impact parameters.

   \item The weak-lensing-deduced mass and SZ distributions are inconsistent with models in which NBDM is not gravitationally dominant. The resulting weak lensing would necessarily arise from the gas and be coincident with the SZ and X-ray signals; this is not what is seen. Further, both lensing and SZ would be more circular than seen due to greater gravitational opposition to the gas expansion perpendicular to the merger axis.

\end{enumerate}

\section*{Acknowledgments}

We thank the anonymous reviewer for very useful comments that have improved the paper, the MRAO staff for their invaluable assistance in the commissioning and operation of AMI which is supported by Cambridge University, and Georgiana Ogrean for providing the Chandra image.
CR is grateful for a STFC Studentship and the support of the Astrophysics Group, Cavendish Laboratory.
MO and YCP acknowledge Research Fellowships from Sidney Sussex College and Trinity College, Cambridge, respectively. We used the COSMOS Shared Memory system at DAMTP, Cambridge University, operated on behalf of the STFC DiRAC HPC Facility, funded by BIS National E-infrastructure capital grant ST/J005673/1 and STFC grants ST/H008586/1, ST/K00333X/1.

\setlength{\labelwidth}{0pt} 

\label{lastpage}
\end{document}